# Thin flexible multi-octave metamaterial absorber for millimetre wavelengths

GIAMPAOLO PISANO,[1,2,*] CHRISTOPHER DUNSCOMBE,[2] PETER HARGRAVE,[2] ALEXEY SHITVOV,[3] CAROLE TUCKER[2]

[1]*Dipartimento di Fisica, Sapienza University of Rome, Rome, Italy*
[2]*School of Physics and Astronomy, Cardiff University, CF24 3AA Cardiff, UK*
[2]*School of Physics and Astronomy, University College London, WC1E 6BT London, UK*
*\*Corresponding author: giampaolo.pisano@uniroma1.it*

**Development of novel radiation-absorbent materials and devices for millimetre and submillimetre astronomy instruments is a research area of significant interest, and with substantial engineering challenges. Alongside low-profile structure and ultra-wideband performance in a wide range of angles of incidence, advanced absorbers in Cosmic Microwave Background (CMB) instruments are aimed at reducing optical systematics, notably instrument polarisation, far beyond previously achievable specifications. This paper presents an innovative design of metamaterial-inspired flat conformable absorber operating in a wide frequency range of 80-400 GHz. The structure comprises a combination of sub-wavelength metal-mesh capacitive and inductive grids and dielectric layers, making use of the magnetic mirror concept for large bandwidth. The overall stack thickness is a quarter of the longest operating wavelength and is close to the theoretical limit stipulated by Rozanov's criterion. The test device is designed to operate at 22.5° incidence. The iterative numerical-experimental design procedure of the new metamaterial absorber is discussed in detail, as well as the practical challenges of its manufacture. A well-established mesh-filter fabrication process has been successfully employed for prototype fabrication, which ensures cryogenic operation of the hot-pressed quasi-optical devices. The final prototype, extensively tested in quasi-optical testbeds using a Fourier-transform spectrometer and a vector network analyser, demonstrated performance closely matching the finite-element analysis simulations, viz., greater than 99% absorbance for both polarisations, with only 0.2% difference, across the frequency band of 80-400 GHz. The angular stability for up to ±10° has been confirmed by simulations. To the best of the authors' knowledge, this is the first successful implementation of a low-profile ultrawideband metamaterial absorber for this frequency range and operating conditions.**

*http://dx.doi.org/10.1364/AO.99.099999*

## 1. INTRODUCTION

The rapid development of millimetre-wave astronomy instrumentation is partly driven by the scientific opportunity provided by Cosmic Microwave Background (CMB) observations. In particular, ongoing CMB polarisation measurements demand unprecedented instrument sensitivity and bandwidth. Proposed millimetre-wave astronomy telescopes feature large multichroic detector arrays operated at sub-Kelvin temperatures. High polarisation sensitivity can be achieved with the aid of a rotating half-wave plate. The corresponding cold optics are designed for maximum optical throughput, wide field of view, and multiple frequency bands. As the cold detector sensitivity approaches theoretical limits and atmospheric systematics become less of an issue for space-borne instruments, low-level instrument systematics have emerged as a crucial requirement for CMB telescopes. These include instrument polarisation effects, notably those associated with the rotating half-wave plate, beam fidelity, and wavefront impairments caused by stray light channelled by spurious reflections along the optical chain. In relation to stray light, there is an ongoing effort to design new absorber material for the optical cavities of the LiteBIRD telescopes [1], which will be deployed in space at the end of the 2020s to carry out an ambitious mission to detect primordial gravitational waves through observations of the CMB B-mode pattern.

Millimetre-wave and infrared absorbers are routinely used in CMB instrument design, e.g. [2], to prevent reflection from the walls and surfaces of the cryostat, pupils, and baffles, as well as other structural elements along the optical path, thus terminating the stray light at the higher temperatures before it reaches the focal plane. Absorbers are designed to meet a range of interrelated, and often conflicting, optical, structural, and thermal requirements, including low reflectivity, wide bandwidth, angular and polarisation independence, small thickness, high mechanical strength, high thermal conductivity at cryogenic temperatures, practical conformability, low weight, and low cost.

Conventional millimetre-wave telescope absorbers include open foam materials loaded with carbon or stainless-steel particles [3]. Tessellated terahertz radiation absorptive material (RAM), developed by Thomas Keating Ltd. [4], is a carbon-loaded polypropylene compound manufactured by injection moulding. It features pyramidal anti-reflection structure on the exterior face and provides low reflectance in 50-1000 GHz frequency range in a wide range of angles of incidence. However, this material is stiff, thick, and heavy, which makes it inconvenient for covering large conformal surfaces inside cylindrical optical cryostats. A modification of such RAM has been reported in [2]. Two-component conductively loaded epoxies, such as thermally conductive epoxy encapsulant Stycast 2850FT by Henkel Corp. [5], are also commonly used as submillimetre-wave absorbers in low-temperature applications, c.f., [6]. A coat of Stycast2850FT usually exhibits ~15% reflectivity, depending on the surface roughness. Epoxy coating can be moulded to reduce reflectivity at lower frequencies, but the process becomes costly. A graphite-loaded epoxy-based moulded pyramidal absorber, constituting a cryogenic thermal source and demonstrating <0.1% reflectance in 75-330 GHz spectral range, was reported in [7], alongside an accurate design model based on geometrical optics analysis. Considerable effort has been dedicated to adopting additive manufacturing techniques for millimetre-wave RAM structures, including 3D printed pyramidal absorber moulds [8], and Hilbert-curve impedance matched structures [9]. Pyramidal tapers of various shapes represent the mainstream technology of broadband absorbers in microwave through infrared optical ranges, c.f., [10], [11]. Conventional open-foam and epoxy based broadband millimetre-wave RAM suffer from drawbacks such as high volume, weight, and low conformability, which affect the thermal budget of the cryogenic optical tubes and complicates absorber installation. An alternative electromagnetic absorber design approach is based on the use of thin multi-layered engineered surfaces, conventionally referred to as planar metamaterials or metasurfaces. The use of the patterned conductive and resistive surfaces with controlled surface impedance, as well as judicious choice of the dielectric interfaces enables robust and cost-effective solutions for millimetre-wave absorbers, as detailed in the following section.

The paper is organized as follows. In Section 2, a review of the planar electromagnetic absorbers is carried out, while the working principle of the new design is detailed in Section 3. Section 4 presents the device's modelling and its initial design. Preliminary fabrication and measurements of the absorber breadboards, aimed to evaluate the actual parameters of the materials and processes, are reported in Section 5. The refined recipe is used to fabricate the final device, and the respective results of its experimental characterization are presented in Section 6. The final conclusions are drawn in Section 7.

## 2. ABSORBING METASURFACES AT MICROWAVE AND MILLIMETRE- WAVE FREQUENCIES

A classic example of the absorbing surface design is the Salisbury screen, [12], described in detail in Section 3, where a resistive sheet is placed at a quarter-wavelength distance from a metallic reflector. Thinner structures can be realized using a high-impedance ground plane (HIGP) instead of the metallic reflector, [13] and [14]. Another type of the device, the Dallenbach absorber, comprises a grounded quarter-wave slab of a lossy dielectric material [15]. Dielectric thickness specified in terms of the wavelength is the major cause of the narrowband performance of the conventional Salisbury and Dallenbach absorbers. The Jaumann absorber [13], comprising multiple alternating layers of resistive sheets and low-density spacers, as well as multi-layer Dallenbach structures allow significant increase of the bandwidth, yet at the expense of the increased volume. Some hybrid broadband structures, combining the features of the three-dimensional and planar multi-layer absorbers, have also been reported in microwave [16], and optical [17] frequency bands. HIGP-based structures appear to be inherently broadband.

Using patterned resistive grids instead of uniform sheets offers additional geometrical degrees of freedom in the design of thin broadband electromagnetic absorbers, [13] and [18]. Planar frequency selective surfaces (FSSs) with engineered surface impedance provide an effective means of reducing the thickness of absorbers. The respective design concept is known as the circuit-analogue absorber (CAA) [19]. The resonant FSS layers could be either resistive or conductive stacked with uniform resistive sheets. Multi-octave bandwidth enhancement can be achieved either with a single-layer FSS comprising a double-periodic array of multi-resonant dipoles or nested resonators [20], or by stacking vertically single-resonant FSS sheets [21], or by combining both approaches [22]. However, design of wideband resonant FSS absorbers proved to be very involved, particularly, with respect to the choice of the resonant elements and harmonic frequencies. Also, such absorbers are inherently limited to a narrow angular range, due to polarisation-dependent response at oblique incidence. Implementing intricate designs of the broadband FSS absorbers in terahertz range is highly challenging technologically.

The use of metasurfaces made of capacitive grids of sub-wavelength square patches, emulating a low-pass filter response, allows one to overcome the limitations of the resonant FSS structures in terms of the bandwidth and angle of incidence. The CAA design paradigm still applies. It has been shown theoretically that a capacitive CAA device, designed using 4 free-standing capacitive grids of resistive patches, could exhibit 20 dB absorption within a ~10:1 frequency range at normal incidence [19]. A large bandwidth with good angular coverage was demonstrated with a structure comprising two lossy capacitive grids, a uniform resistive sheet and 5 dielectric layers with low and high refractive indexes. The simulated structure exhibited a 20 dB absorption over a 7.5:1 bandwidth for up to 45° angle of incidence for both TE and TM polarisations. Both structures reported in [19] feature nearly optimum thickness for the given bandwidth, according to the Rozanov's criterion derived in [23], although their performance was verified by simulations only. A recent microwave metamaterial absorber structure in [24], designed with circuit modelling, comprises three layers of square resistive patches interspersed with dielectric layers and demonstrates an ultra-wideband and wide-angle absorption response from 4.73 to 39.04 GHz (i.e., 8.25:1 band ratio) in simulations, which was partly confirmed by measurements within a reduced frequency band. The device thickness appeared to be a factor of 0.15 of the longest operating wavelength. A great variety of circuit-model designs of microwave absorbers aimed at achieving the optimum performance in terms of the bandwidth, thickness, and reflectivity, have been reported in the literature, predominantly by simulations only, but to the best of our knowledge there has been only a rudimentary attempt to implement such structures in the millimetre-wave through terahertz range, not to mention their experimental demonstrations. It must be noted that such implementation does not reduce to a simple scaling of the geometry, because of specific behaviour of materials in the millimetre-wave range under specific operation conditions, as further discussed below.

The use of exotic metamaterial phenomena arising from concerted electric and magnetic polarisabilities of sub-wavelength scatterers allow even greater freedom in the design of millimetre through optical range absorbers. Although widely viewed as inherently narrowband and lossy, advanced dispersion-engineered metamaterial and metasurface structures feature increasingly broader performance. A three-dimensional honeycomb-like metamaterial absorber reported in [25] demonstrates in simulation a 90%-absorption bandwidth from 50 to 460 GHz (1:9.2 bandwidth ratio), as well as polarisation and angle independence. A concept of Huygens' metasurfaces, featuring inherent unidirectional scattering, has been employed to design a perfect metamaterial absorber [26], although at a single frequency. A conceptual multi-band absorber for infrared and optical bands proposed in [27] exploits interference phenomena in stacked layers of epsilon-near-zero metamaterial and high-permittivity dielectric. It has been shown that bandwidth enhancement could be observed in plasmonic metamaterial absorbers [28].

In our opinion, despite the high volume of research on dispersion-engineered metamaterial absorbers, such devices still remain relatively narrowband and lossy, see e.g. [29] for relevant discussion, and less practical from the points of view of design and manufacture, as compared to the FSS and capacitive-grid absorbers. In addition, most of the research is based solely on simulations, with very few frequency-downscaled prototypes having ever been manufactured and tested.

This paper presents the theory, design, fabrication, and experimental characterization of a novel millimeter-wave absorber based on a very simple working principle. It is based on the use of a broadband magnetic mirror, rather than a metallic mirror, which can be built with dielectric layers and with just one resistive sheet [30]. The higher the number of layers of the absorber, the larger the bandwidth. For the sake of demonstration, we set our requirement to those we use for CMB instruments, meaning reflectivity at -20dB level (i.e. 99% absorption) over a large frequency bandwidth (~80 to 440 GHz, i.e. a 5:1 ratio), independence between S and P polarisations (perpendicular and orthogonal to the radiation plane of incidence), and high performance within a wide range of incidence angles (e.g. 0°-50°). We also aimed to develop a thin and flexible profile, compatible in terms of materials and manufacturing processes with metal-mesh filters [31], flat lenses [32], and metamaterial half-wave plates [33], that we have developed in the past. The new absorber comprises four dielectric layers and one resistive mesh. The refractive indices of two layers were artificially synthesised by embedding sub-wavelength capacitive meshes within polypropylene. The device was designed using the same formalism employed we used to develop an artificial magnetic conductor [30] based on the mesh filters technology. The new device, which has never been demonstrated before, successfully achieved these performance requirements.

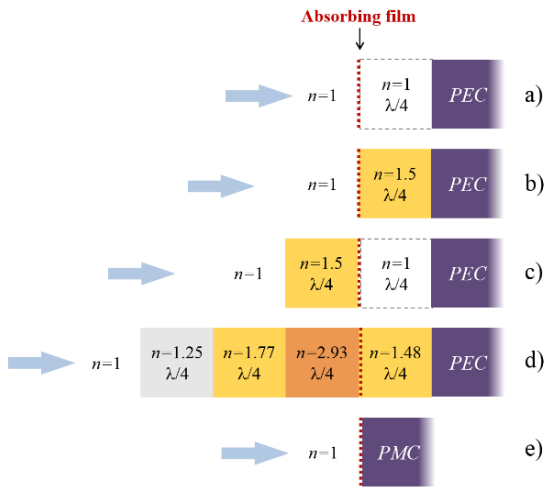

Fig. 1. Sketch of different types of absorbers based on an absorbing film: a) Salisbury screen (air-gap); b) Salisbury screen on substrate; c) Salisbury screen on substrate and air-gap; d) realization of a magnetic mirror absorber; e) ideal magnetic mirror based absorber.

## 3. MESH-ABSORBER SURFACE WORKING PRINCIPLE

Here we describe the working principle of the new absorber which implies a transition from the conventional means of absorbing radiation using a perfect electric conductor as backshort, to the idea to use a perfect magnetic conductor in its place.

### A. Salisbury Screen

The simplest design for an effective absorbing surface is the Salisbury screen [12], [13]. It consists of an absorbing sheet located at a quarter of the wavelength distance $\lambda_0/4$ from a metallic mirror (see Fig. 1a). The wavelength $\lambda_0$ defines the frequency $\nu_0$ where the absorption is maximum. The radiation passing through the sheet will be in phase with that bouncing back from the mirror. This happens because the latter gains a phase-factor $\pi$ in the metal reflection and another factor $\pi$ for the half-wavelength extra path back and forth, for a total of $2\pi$. The standing wave so created has its electric field maximum exactly where the absorbing sheet is. The surface impedance of the sheet is equal to the free-space impedance and provides unitary absorption and minimum reflection at the central frequency $\nu_0 = c/\lambda_0$, where $c$ is the speed of light. These structures can be easily modelled using transmission-line or propagation-matrix codes. In this work we have chosen our central frequency to be $\nu_0 = 240$ GHz. The frequency dependent absorption and reflection coefficients for a Salisbury screen are shown in Fig. 2, curves labelled a). The averaged absorption (and reflection) coefficient across a 5:1 bandwidth around $\nu_0$, i.e., from 80 GHz to 400 GHz, are respectively $A = 87.8\%$ and $R = -9.1$dB. This is obtained by setting the absorber surface impedance equal to the free space impedance, $Z_{abs} = 377\,\Omega/\square$.

We note that, if the absorbing sheet has a frequency-independent impedance, the equiphase condition will be also satisfied at higher frequencies, those for which the absorber-backshort distance is an odd multiple of their quarter wavelength, i.e., when $\lambda_0/4 = (2m+1)\,\lambda_m/4$, where $m = 0, 1,...$ The result is a periodic absorption (and reflection) behavior with maxima (and minima) at harmonic frequencies $\nu_1 = 3\nu_0$, $\nu_2 = 5\nu_0$, etc.

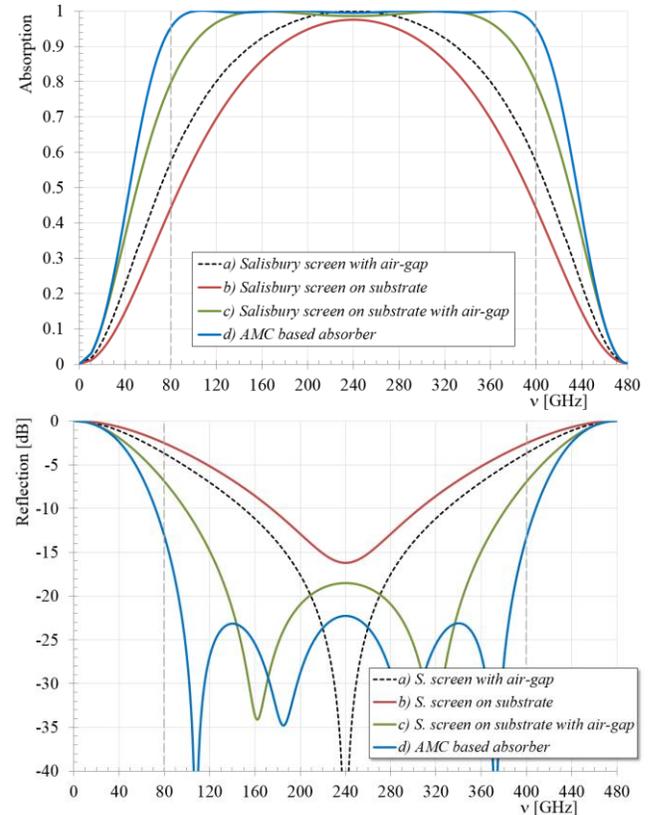

Fig. 2. Absorption and reflection coefficients vs frequency for the different absorbers a) to d) sketched in Fig. 1.

### B. Salisbury Screens on substrates

The goal of our work is twofold: i) to design a robust absorbing device supported by dielectric substrates that could be manufactured using the mesh-filters technology; 2) to obtain high absorption over very large bandwidths.

The first attempt would be to study an embedded version of the Salisbury screen. It could be made by replacing the $\lambda_0/4$ air-gap with a quarter-wavelength dielectric layer ($\lambda_0/4n$) and by depositing the absorber on its external surface (Fig. 1b). This would lead to a device much more robust than the previous one. However, by using for example polypropylene as a substrate (refractive index $n \cong 1.5$), the optimized design (with $Z_{abs}$ = 273 $\Omega/\square$) would decrease the average absorption within the 5:1 band down to $A$ = 80.9% ($R$ = -7.2 dB), as shown in Fig. 2, curves labelled b). This is because some additional out-of-phase reflection is now added at the free-space to dielectric boundary, where the absorber is. Indeed, a mismatch from a low to high index implies a negative reflection coefficient.

We can look at a different configuration where the substrate and the absorbing film are now reversed compared to the incoming radiation direction, i.e., we keep the absorbing film at the same distance from the metal backshort but we reverse the position of its substrate, as shown in Fig. 1c, The optimized absorption curve (obtained with $Z_{abs} \cong 213 \, \Omega/\square$), changes dramatically and its 5:1 band average boosts up to $A$ = 96.8% ($R$ = -15.0 dB); see curves labelled c) in Fig. 2. This is because some additional in-phase reflection is now added at the dielectric to free-space boundary (high to low index), where the absorber is. The effect that we are describing is at the roots of the main idea of this work, i.e., forcing most of the radiation to be reflected in phase where the absorbing surface is. This can be achieved by using a magnetic mirror, as described in the next section.

We note that the latter configuration, although showing good performance, would be difficult to manufacture because it would require a structure able to maintain the air-gap distance constant over large surfaces. It would be even more challenging to develop it if the absorbing surface is not meant to be flat.

### C. AMC-based absorber

Magnetic mirrors, or Artificial Magnetic Conductors (AMCs), are surfaces designed to mimic the behaviour of ideal Perfect Magnetic Conductors (PMCs). These ideal materials reflect 100% of the incident radiation, they provide a null phase-shift and can be modelled as surfaces exhibiting infinite impedance. This contrasts with normal Perfect Electric Conductors (PECs), which provide a $\pi$ phase-shift and can be modelled with a null surface impedance.

As we have seen in the previous section, the quarter-wavelength air or dielectric gap is the main factor affecting the bandwidth and a natural way to improve absorption is to have in-phase reflection right at the plane of the absorber. By definition, the radiation reflected off a magnetic mirror is in phase with the incident one and an absorber located right on its surface should greatly increase its efficiency. This means that the maximum of the electric fields is not anymore localized at a certain $\lambda_0/4$ distance from the mirror-backshort but directly on its surface. This does not impose any geometrical constraint or frequency dependence to the system. Such an ideal system, sketched in Fig. 1e, would absorb radiation at all frequencies, i.e., it would have an infinite bandwidth.

There are many ways to design magnetic mirrors (or AMCs). Here we base our work on a device developed using the mesh-filter technology that led to multi-octave bandwidth operation [30]. The working principle of this magnetic mirror, part of the device sketched in Fig. 1d, is based on the null phase-shift obtained in the reflection occurring at a high-to-low index boundary. The higher the index difference, the higher the reflection coefficient. For this reason, a graded index section is used at the input of the device to drive the radiation adiabatically into a high index medium. A sudden jump into a lower index medium provides a high reflection coefficient with null phase shift, over large bandwidths. A backshort located at a quarter wavelength in that medium defines the central frequency of operation.

The broadband mesh-absorber presented here is made with a magnetic mirror similar to the one described above with an embedded absorber film located at the high-to-low index interface (Fig. 1d).

## 4. MODELLING AND INITIAL DESIGN

The absorber presented in this work is based on the mesh-filter technology. In this section we briefly describe this technology, how we model the device and its initial design, which will not include the details of the absorbing film, discussed in Sec. 5.

### A. Mesh-filter technology

The mesh absorber was manufactured using the processes developed for mesh-filters [31]. This has been employed not only to realise filters, used in many astronomical instruments operating at millimetre and sub-millimetre waves, but also to develop novel devices such as flat mesh lenses [32] and mesh half-wave plates [33]. The devices are manufactured by embedding metal grids within polypropylene layers, and by hot-pressing them all into a single homogenous polypropylene matrix where the metal grids remain suspended within it. In this specific application the grids are not designed to interfere, like in above mentioned devices; they are instead closed-packed to create the artificial dielectrics [34] required for the graded index medium of the absorber.

A sketch of the ideal mesh-absorber made with homogeneous materials and its actual realization employing mesh-filter technology are reported respectively in Fig. 3a and Fig. 3b. The ideal device requires four quarter wavelength layers: the first three with increasing refractive indices ($n_1$ = 1.25, $n_2$ = 1.77 and $n_3$ = 2.93) and the last one with a low intermediate index ($n_4$ = 1.48). The absorbing film, with optimized surface impedance $Z_{abs} \cong 103 \, \Omega/\square$, is located between the third and the fourth layers. The metal backshort is at the end of the stack. The absorption curve is now very close to 1 across the 5:1 bandwidth around $\nu_0$, see curves labelled d) in Fig. 2, and its average value is boosted up to $A$ = 99.5% ($R$ = -23dB).

The materials generally employed to build mesh-devices are: i) porous PTFE ($n_{pPTFE} \cong 1.25$), used as anti-reflection (AR) coating; ii) polypropylene ($n_{PP} \cong 1.48$), used as the substrate supporting the embedded metal grids. These two materials can be used respectively for the first and the fourth layer of the mesh-absorber, so that $n_1 = n_{pPTFE}$ and $n_4 = n_{PP}$. However, the indices $n_2$ and $n_3$ can be artificially synthesized by embedding close-packed copper grids into polypropylene, following a method described elsewhere [34]. The two quarter-wavelength sections will require pairs of embedded capacitive grids, i.e., periodic square patches, with different filling factors. The capacitive grids have a period of $g$ = 100μm and this size allows the artificial media to be 'seen' by the radiation as homogeneous dielectrics, with indices very close to $n_2$ and $n_3$, across a large bandwidth (0-500 GHz) [30].

The material used for the absorbing film is bismuth. The film will be required to have a specific pattern and surface impedance; this will be discussed in detail in Sec. 5. The metal backshort is realized with a 400 nm copper deposition.

The final device, apart from the AR-coating layer, is completely made with polypropylene with the four capacitive grids, the absorber and the backshort either embedded or deposited on it (Fig. 3b). The total thickness of the device is ~750 μm, equivalent to ~0.6 λ and ~0.2 λ respectively at the central and lowest operational frequencies.

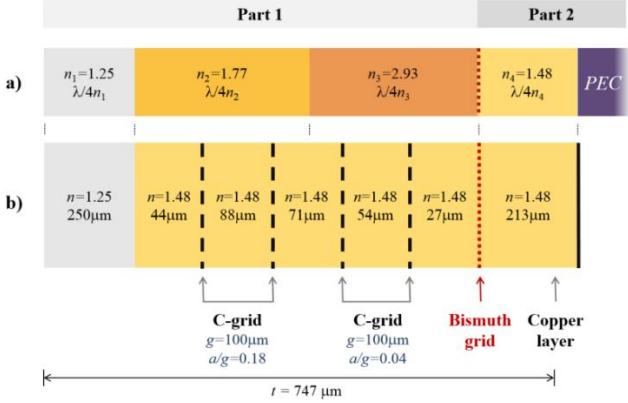

Fig. 3. a) Sketch of the magnetic mirror absorber made with homogeneous materials; b) Sketch of the mesh-absorber realization using copper grids and the bismuth layer embedded into polypropylene.

### B. Finite-element modelling

As mentioned in Sec. 3, the performance of the devices sketched in Fig.1 have been computed using a propagation matrix code. In this case the absorbing films were modelled as shunt admittances, the various layers as homogenous media and the radiation assumed to be at normal incidence. The actual mesh-absorber device is made with homogeneous layers but also with metamaterials which can be accurately simulated using finite-element analysis (FEA). We have used the commercial software Ansys HFSS [35] for our preliminary modelling, detailed design and parameter retrieval of the final manufactured device.

The HFSS models consist of unit cells with periodic boundaries that mimic infinite arrays (Fig. 4). The boundaries are of the 'master & slave' type to allow the simulation of radiation at any angle of incidence and for both types of polarisations (S or P). The porous PTFE and the polypropylene substrates are modelled as homogenous dielectric materials (the latter with a frequency dependent loss tangent), the metal grids as thin copper patches with finite conductivity. The absorbers are modelled in two different ways: a) as homogeneous surfaces with associated impedance; b) as patterned surfaces with associated surface impedance. The reason for using these two types of models will be clarified later.

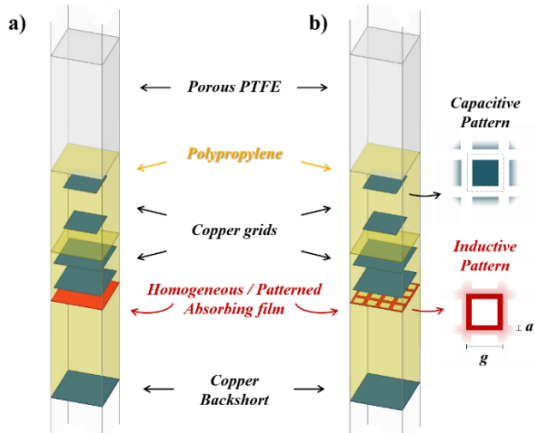

Fig. 4. Mesh-absorber HFSS models: a) homogeneous absorbing surface; b) patterned (inductive) absorbing surface.

In designing an absorbing device, in addition to low reflectivity and large bandwidth, there could be also requirements in terms of incidence angles. Depending on the application, the device might be required to work off-axis, or to maintain high performance over a wide range of angles. In these cases, we expect the absorption to vary because the radiation will travel different path lengths through the layers of the absorber. However, the absorber can be designed to have maximum performance at a specific angle $\theta$, and this could correspond to the average angle within the above range. In our case, to prove the device working principle we have chosen $\theta = 22.5°$, which corresponds to the reflection angle of our testing setups (see Sec. 6). We note that working off-axis implies the distinction between S and P polarisations, and the absorber will then need to be efficient in both configurations.

The initial design was simulated using a model of the type shown in Fig. 4a, i.e., with a homogeneous absorbing surface between the third and the fourth layers. The surface impedance was varied in the range 95 - 115 Ω/□, the incidence angle was set to $\theta = 22.5°$ and both S and P polarisation absorption coefficients evaluated. The results of these simulations are reported in Fig. 5 and Fig. 6 for the S and P polarisations, respectively. The best averaged off-axis absorptions across the 5:1 bandwidth (80-400 GHz) are achieved by choosing values of surface impedance $Z_{abs\_22.5\_S} = 105$ Ω/□ and $Z_{abs\_22.5\_P} = 100$ Ω/□, respectively for the S and P polarisations. The average of these values is close to the one obtained in the on-axis case, i.e., $Z_{abs\_0} \cong 103$ Ω/□. By choosing the latter as an off-axis trade-off value, the averaged absorption appears to be almost polarisation independent, at -21.6 dB level for both S and P polarisations. We note that the off-axis operation slightly shifts the operating band to higher frequencies (Fig. 5 and Fig. 6).

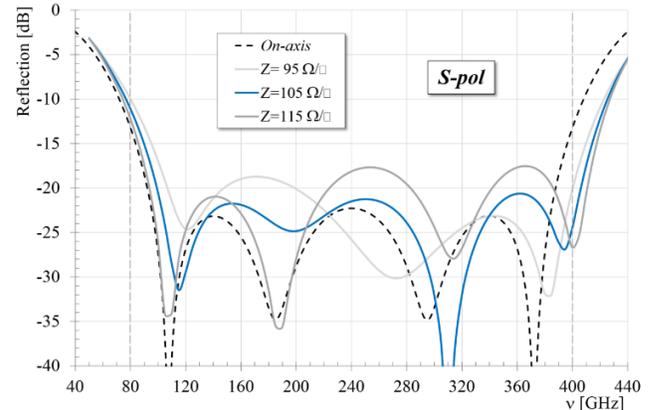

Fig. 5. S polarisation reflection coefficient vs. frequency for 22.5° incidence angle and different impedances for the absorber.

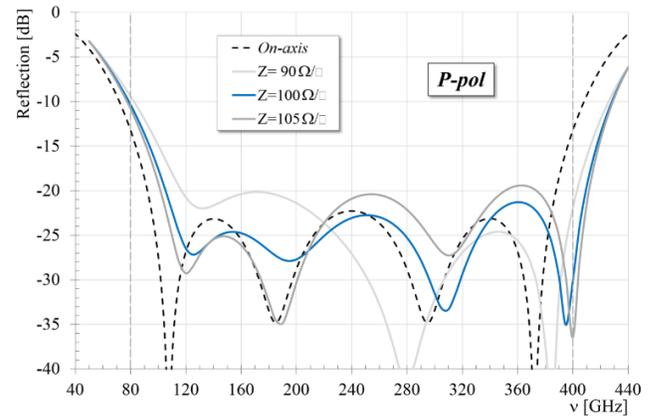

Fig. 6. P polarisation reflection coefficient vs. frequency for 22.5° incidence angle and different impedances for the absorber.

## 5. MANUFACTURE AND DESIGN FINE-TUNING

The manufacture of the mesh-absorber followed different steps: a) manufacture and tests of the graded index section (Part 1); b) R&D on uniform resistive films; c) R&D on patterned resistive films; d) heat bonding test with a dummy absorber device; e) manufacture of the backshort and the final assembly.

### A. Part 1 - Graded index

The graded index part of the absorber, Part 1 in Fig. 3, was the first to be manufactured using the standard mesh-filter processes. It consisted of three quarter-wavelength layers: the first made with pPTFE ($n_1$), the second and the third with polypropylene embedded capacitive meshes designed with their geometry and spacing to achieve the effective refractive indices of $n_2$ and $n_3$. The measured thickness of the assembly was close to the nominal value within the measurement error (~2 μm). Transmission measurements were performed on-axis using the FTS to check the performance of this part of the absorber (see Fig. 7). A finite-element model of Part 1 was built to compare its predictions with the measured data. The shape of the transmission curve of Part 1 is not really relevant but indeed useful to extract more accurate values of the various parameters. The measured data, up to 600 GHz, goes beyond the frequency range of the device. The transmission peak around 480 GHz is used to fit the data with higher accuracy. A four-parameter optimisation of the finite-element model was run across a discrete number of frequency points to fit the measured data. The parameters were the pPTFE and polypropylene refractive indices and the $a/g$ parameters of the two pairs of capacitive grids (see Fig. 4). The overall thickness of Part 1 was set to the nominal one and the grid period ($g = 100$ μm) was not varied. The copper conductivity was assumed to have the standard value we use for these grids, $\sigma_{Cu}=4 \times 10^7$ S/m. The fit procedure led to the following values: $n_{pPTFE}=1.23$, $n_{PP}=1.48$, $(a/g)_1=0.186$ and $(a/g)_2=0.045$. The refractive indices fall in the expected ranges of variability related to the bonding processes. The slightly higher fitted values of $a/g$, as compared to the design values, imply over-etching of the capacitive grids: the sides of the square patches resulted to be smaller by ~1 μm in both pairs. The values of all the above parameters, either measured or estimated, will be used from now on to model this part of the device and to optimise other parts, as discussed in the following sections.

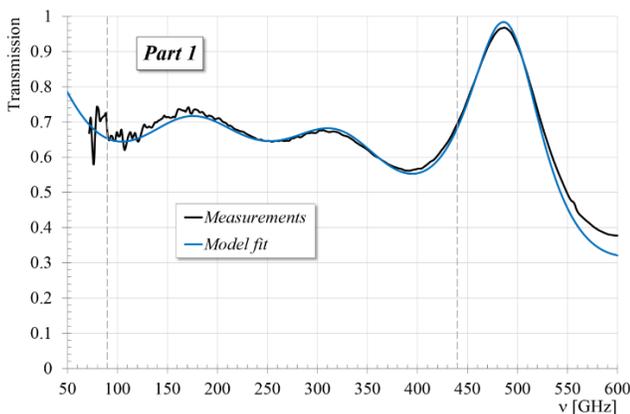

Fig. 7. Part 1 on-axis transmission measured and finite-element best fit model obtained by varying the pPTFE and PP refractive indices and the grid geometries.

### B. Homogeneous absorbing films

One way to realise a resistive sheet consists in evaporating a thin film of metal on a polymer substrate. If the film thickness is well below the skin depth $\delta_S$ at all the frequencies of operation, the radiation going through the film will interact with the resistive layer and some of its power will be dissipated across it. For a given resistivity $\rho$ and thickness $t$, the film surface impedance equals $Z_S=\rho/t$. If $t << \delta_S$ then the surface impedance can be considered constant with frequency.

In the case of an ideal free-standing resistive film, the transmission, absorption and reflection coefficients as a function of the surface impedance can be easily computed using a transmission line circuit. From these curves, reported in Fig. 8, we note that by measuring the transmission coefficient of an absorbing film, we can infer immediately its surface impedance.

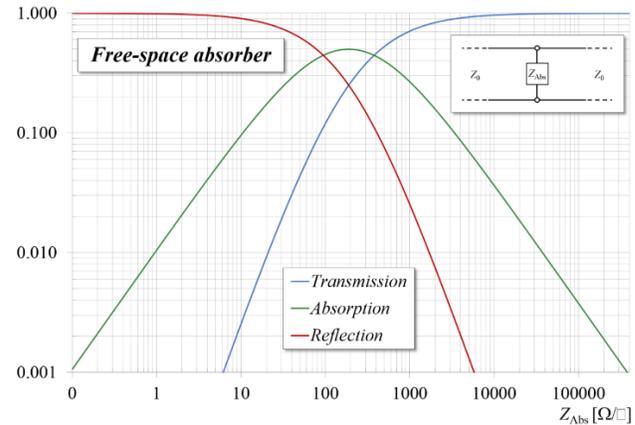

Fig. 8. Ideal free-space absorber transmission, absorption and reflection coefficients as a function of its surface impedance. The absorber is modelled as an infinite surface with no thickness.

One of the outcomes of the preliminary design is that the absorbing film of our device needs to have a surface impedance of the order of $Z_S \approx 103$ Ω/□. Materials such as copper or gold cannot be used, because their high electrical conductivity would imply film thicknesses less than 1 nm. What is required is a metal which much lower conductivity, at least two orders of magnitudes below, that could be also evaporated on a polymer substrate and processed using the mesh-filter techniques. The conductivity of a thin film of evaporated bismuth depends on its thickness and varies typically in the range $\sigma_{Bi} \sim (0.35 \div 1.92) \times 10^5$ S/m for thicknesses $t = 14 \div 220$ nm [36]. Using these values as starting point, the targeted $Z_S \sim 103$ Ω/□ could be achieved with a bismuth film roughly 70nm thick.

Before processing a specific absorbing film, it was required to carry out some bismuth evaporation tests on the polymer substrates normally used in mesh devices: Mylar and polypropylene. The bismuth deposition was processed via thermal evaporation in a $10^{-6}$ mbar vacuum. Achieving reproducibility in the evaporation processes was one of the most critical factors. The first samples were tested in transmission using an FTS test-bed covering the frequency range of 150-600 GHz. Examples of these measurements are shown in Fig. 9. Having verified the very flat response with frequency of the films, the samples manufactured subsequently were tested using a VNA test-bed across a narrower frequency range, i.e. 160-260 GHz.

The first sample, Bi-1, was a 45 nm Bi film on a 0.9 μm thick Mylar substrate, whereas the second, Bi-2, - a 450 nm Bi film on a 4 μm thick polypropylene. The Bi thickness was measured using a Quartz crystal monitor which was calibrated using a surface profilometer (error of ±5%). These samples were useful to validate both the modelling tools and processing.

Bi-1 and Bi-2 samples showed constant transmissions of the order of $T_{Bi-1} \cong 0.252$ and $T_{Bi-2} \cong 0.003$, respectively. The curves in Fig. 8, although not including the small effects of the thin substrates, can be used to infer the associated surface impedances with a good degree of accuracy (at a

1% level): $Z_{Bi-1} \sim 189\,\Omega/\square$ and $Z_{Bi-2} \sim 10\,\Omega/\square$. Simplified versions of the HFSS model shown in Fig. 4a have been used to infer the conductivity of bismuth. These models consisted of a dielectric substrate and a thin metal layer with variable conductivity. Optimisation fits were run to match the transmission coefficients of the model and measured data, yielding the following conductivity values: $\sigma_{Bi-1} = 1.2 \times 10^5$ S/m and $\sigma_{Bi-2} = 2.2 \times 10^5$ S/m. These values are consistent (although outside their measured range) with those reported in [36].

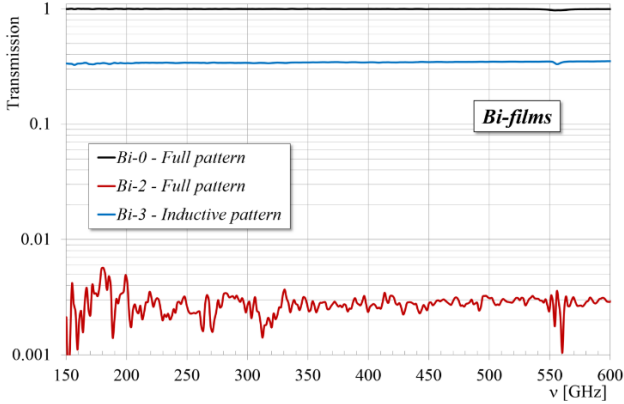

Fig. 9. Transmission measurements of the bismuth film samples.

### C. Patterned absorbing films

A homogeneous absorbing film with the target impedance $Z_S \sim 103\,\Omega/\square$ could be, in principle, processed. However, the final device, as sketched in Fig. 3b, requires the Bi-film to be sandwiched between polypropylene layers, and the associated heat-bonding process does not work in the presence of uniform metal layers. For this reason, the Bi-film needs to be patterned as an inductive grid, e.g., with square holes in it. This would allow the polypropylene layers on either side of the film to penetrate the sheet and bond during the hot-pressing process. We used an inductive pattern with a period $g=25\,\mu m$ and the standard $a/g=0.14$ (see Fig. 4b).

For a given thickness, the surface impedance of a patterned layer of bismuth will be higher than for the uniform film, due to its dilution across the unit area. This means that the target surface impedance will be achieved with a thickness which is larger than the one estimated earlier (70nm). The required thickness was estimated using another HFSS model (similar to the previous one but including a patterned bismuth layer) and resulted to be $t \cong 180$ nm.

Again, before proceeding to the final film evaporation, the above process required some development. The inductive pattern was obtained using a lift-off process, rather than the standard etching process on a homogenous layer of bismuth. This was more suitable due to the materials, dimensions and thicknesses involved. However, the lift-off process provided directly the final inductive pattern without giving the possibility to test first the uniform layer.

Two patterned film samples were processed, Bi-3 and Bi4, respectively with a 90nm and a 270nm bismuth layer on 9 μm thick PP substrates. FTS and VNA transmission measurements provided averaged transmissions of $T_{Bi-3} \cong 0.340$ and $T_{Bi-4} \cong 0.050$, corresponding to the effective surface impedances of $Z_{Bi-3} \sim 263\,\Omega/\square$ and $Z_{Bi-4} \sim 54\,\Omega/\square$. Using the HFSS models and running optimisations to fit the data, it was possible to infer the values of the thin film conductivities which resulted to be: $\sigma_{Bi-3}=1.4 \times 10^5$ S/m and $\sigma_{Bi-4}=2.3 \times 10^5$ S/m.

Given the success with the previous processing, the final absorbing film with the targeted $Z_S \sim 103\,\Omega/\square$ could be now manufactured. The patterned sample Bi-5 had a 175 nm thick layer of bismuth on a 9 μm thick PP substrate. Its averaged transmission $T_{Bi-5} \cong 0.122$ implied an equivalent surface impedance $Z_{Bi-5} \sim 101\,\Omega/\square$, not far from the goal, and a film conductivity $\sigma_{Bi-5}=1.9 \times 10^5$ S/m, close to what could be extrapolated from [36], i.e., $\sigma_{Bi}(175\,nm) \cong 1.7 \times 10^5$ S/m.

Table 1. Bismuth absorbing films characteristics

| Film | $t$ | Pattern | $T$ | $Z_S$ | $\sigma_{HFSS}$ |
| [#] | [nm] | | | [$\Omega/\square$] | [S/m] |
|---|---|---|---|---|---|
| Bi-1 | 45 | Full | 0.252 | 189 | 1.2E+5 |
| Bi-2 | 450 | Full | 0.003 | 10 | 2.2E+5 |
| Bi-3 | 90 | I25/0.14 | 0.340 | 263 | 1.4E+5 |
| Bi-4 | 270 | I25/0.14 | 0.050 | 54 | 2.3E+5 |
| Bi-5 | 175 | I25/0.14 | 0.122 | 101 | 1.9E+5 |

### D. Bonding process test with dummy PP layers and absorbing film

All the constituent parts of the mesh-absorber were ready to be assembled. The final process, required to be checked, was the bonding of the patterned bismuth film between PP layers, which had never been done before. For this purpose, the patterned Bi-3 sample was sandwiched and successfully hot-pressed between two layers of PP with thicknesses equal to those of the final device, i.e., 284μm and 213μm.

VNA transmission measurements of this 'inefficient' dummy absorber were conducted to check for any potential variations of the surface impedance during the bonding process (Fig. 10). Two HFSS models were built to simulate the dummy sandwich with either the uniform or patterned resistive films. Both were used to fit the experimental data by varying the substrate's refractive index, equivalent surface impedance (first model), and conductivity of the bismuth thick pattern (second model). The optimisation yielded almost indistinguishable transmission curves and the following values of the film parameters: $Z_{Bi3\_Dummy} \sim 233\,\Omega/\square$ and $\sigma_{Bi-3\_Dummy} \sim 1.6 \times 10^5$ S/m. These results imply a reduction of ~11% of the original impedance as a consequence of the bonding process.

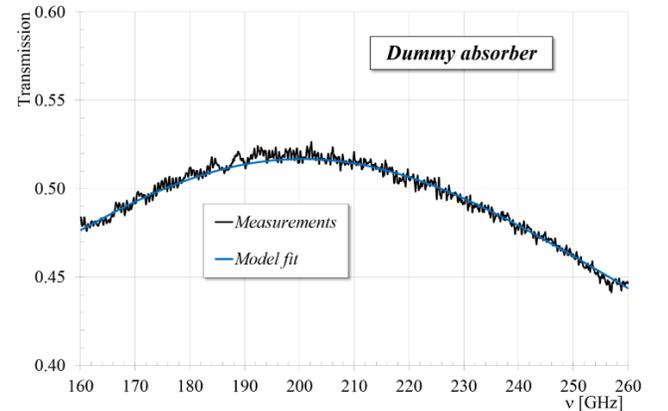

Fig. 10. Transmission measurements of the dummy absorber together with its best model fit obtained by varying the substrate refractive index and bismuth conductivity.

### E. Part 2 and the final assembly

The results provided by the dummy bonding implied a potential reduction of the surface impedance of the sample Bi-5 when bonded within the final device. An HFSS model of the complete absorber, including the patterned Bi-5 film with the nominal ($Z_{Bi-5} \sim 103\,\Omega/\square$) and reduced ($Z_{Bi-5\_red} \sim 91\,\Omega/\square$) values of surface impedance were run for both polarisations. The overall absorption averaged over the 80–400 GHz band appeared to remain below -20dB for both S and P polarisations, thus allowing us to proceed with the assembly of the final mesh-absorber.

The manufacture of Part 2 was straightforward, because this part consisted of bonded layers of PP with uniform copper evaporation on one side (Fig. 3b). The whole device was eventually manufactured by stacking and hot-pressing Part 1, Bi-5, and Part 2. Photographs of the front and back sides of the final mesh-absorber are shown in Fig. 11. An example of the finite-element simulations of the mesh-absorber, run at 250 GHz, are shown in Fig. 12.

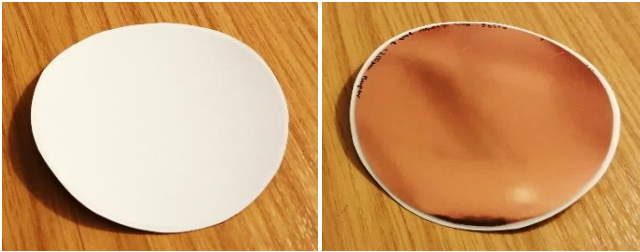

Fig. 11. Photographs of the mesh absorber taken from the AR-coating side (white) and from its copper backshort side. The device has a diameter of 100mm, it is ~750μm thick, and it is mechanically flexible.

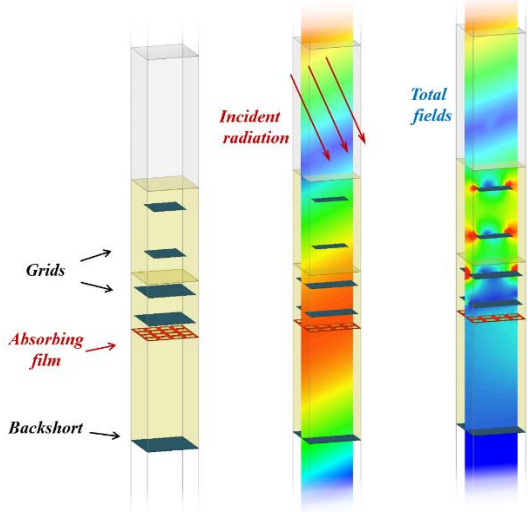

Fig. 12. Finite-element simulation showing the absorption of the electromagnetic field at 250 GHz, with light incident at 22.5° and both S and P polarisations present.

## 6. EXPERIMENTAL CHARACTERISATION

Here we briefly describe the experimental setups used for the various tests and the detailed characterisation of the final mesh-absorber.

### A. VNA and FTS experimental setups

The breadboard samples and the final device were tested using two different testbeds: a Fourier transform spectrometer of the Martin-Puplett type, operating in the 50-600 GHz frequency range by means of a cryogenically cooled bolometer detector, and a Rohde & Schwartz ZVA67 vector-network analyser operating from 75 to 330 GHz by means of standalone frequency extenders. As mentioned earlier, the FTS was initially used to measure the transmission coefficients of the first Bi-samples and verify their broadband flat response. The VNA was then used to extract the surface impedances of the other Bi-samples over narrower frequency ranges and to quantify the Bi conductivity changes in the dummy absorber. The FTS was used for the transmission measurements of Part 1 and for the full-band reflection measurements of the final absorber. The relevant experimental setups have been described in detail elsewhere: the VNA transmission measurements in [37], the FTS transmission measurements in [33], and the FTS reflection measurements in [38].

We notice that the sample flatness proved to be a crucial factor of the data quality in reflection measurements. In the tests, the absorber device was clamped in a holder and placed inside a mount behind an aperture. In the test, the holder was rotated about the optical axis and the small deviations of the received signal provided evidence of suitable flatness.

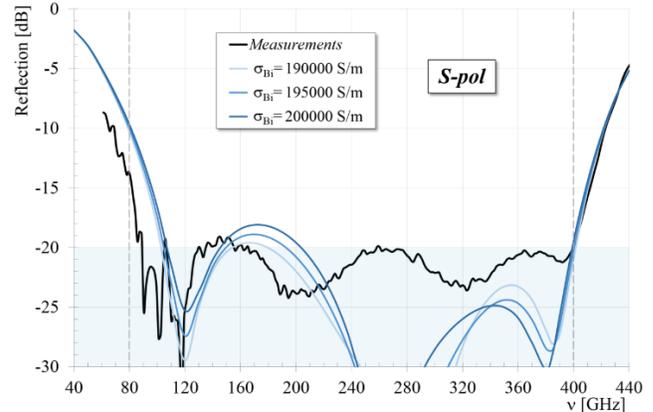

Fig. 13. Mesh-absorber simulated and measured reflection coefficient versus frequency for the S polarisation at 22.5° incidence angle.

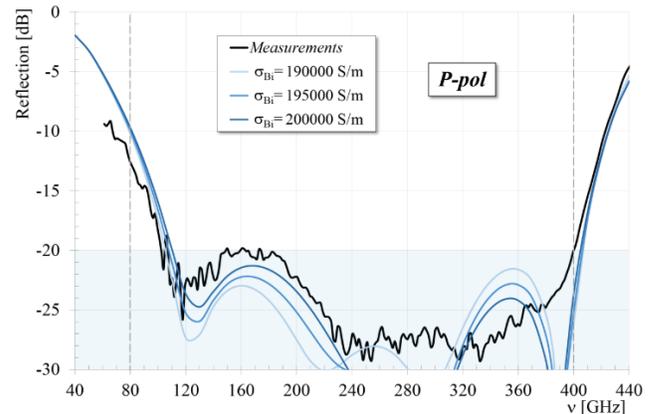

Fig. 14. Mesh-absorber simulated and measured reflection coefficient versus frequency for the P polarisation at 22.5° incidence angle.

### B. Absorber tests and results

The mesh absorber was tested with the FTS at 22.5° incidence angle with both S and P polarisations. The measured data are reported in Fig. 13 and Fig. 14. A very good performance of the device across a wide frequency band is immediately evidenced, for both polarisations. Quantitatively, the measured absorption coefficients averaged in the 80-400 GHz (a 5:1 bandwidth) frequency range, for both polarisations, resulted to be $A_{22.5\_S}$ = 99.2% ($R_{22.5\_S}$ = -21.2 dB) and $A_{22.5\_P}$ = 99.4% ($R_{22.5\_P}$ = -22.0 dB), respectively. These results meet the requirement that was set at the beginning, i.e., absorption to be ≥ 99% (reflection ≤ -20dB). Also, the differential absorption between S and P polarisations resulted to be very small, at 0.2% level.

Although the design and development procedure successfully led to the desired performance, we have tried to fit the final data with an HFSS model of the type b) of Fig.4. Within the many parameters required to run the model, some were identified and fixed previously: the copper conductivity, the device thickness (directly measured) and the capacitive grid geometries $(a/g)_{1,2}$ (extracted by testing Part 1). The remaining parameters, the PP and pPTFE refractive indices as well as the bismuth conductivity, could still change during the final bonding, and we left them vary in the final fit. Changes in $n_{PTFE}$ provided negligible variations in the performance, whereas, as observed in the past [33], a systematic increase of $n_{PP}$ shifted the overall curve towards low frequencies. In addition, as learned by testing the dummy absorber, we also expected an increase of $\sigma_{Bi-5}$. The models that better reproduced the measured data are shown in Fig. 13 and Fig. 14. The required values of the refractive indices were $n_{pPTFE} = 1.23$ and $n_{PP} = 1.51$. We reported the model results for three increasing values of the bismuth conductivity, starting from the initial one estimated before the bonding, i.e., $\sigma_{Bi-5} = 1.9 - 2.0 \times 10^5$ S/m.

There is a good agreement between the simulations and data for both polarisations down to a -20 dB level. Below that, down to the noise floor of the FTS, at roughly -30 dB, there are variations, especially in the S polarisation case. These variations are actually not important, because they are well below the requirement we set. In any case, we have tried to fit the data by going back and varying also the parameters that were initially fixed, but we could not fit all the features. Our explanation is that the grids, when stacked together to manufacture the device, were not necessarily aligned. The capacitive grids were probably displaced (with the square patches not aligned as in Fig. 4) and rotated with respect to each other. An HFSS model with such arbitrary rotations cannot be built using periodic boundaries and the associated small changes in the performance could not be replicated. This observation suggests that future devices should be built with aligned grids.

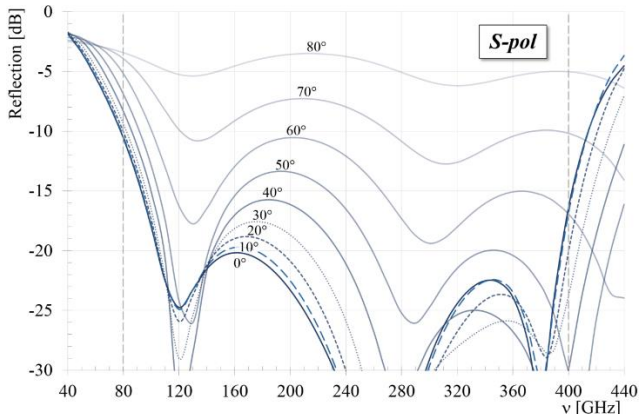

Fig. 15. Mesh-absorber simulated spectral response for different incidence angles for the S polarisation.

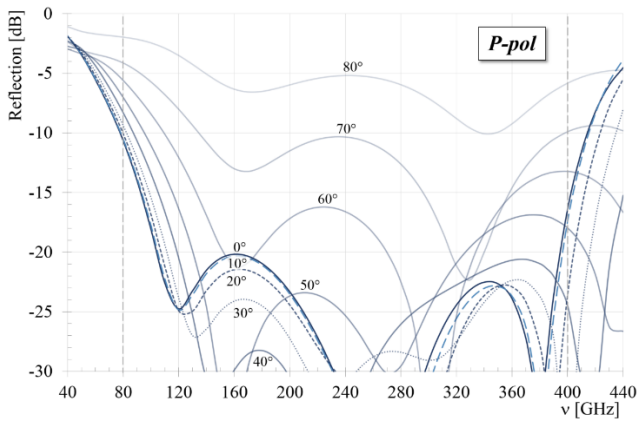

Fig. 16. Mesh-absorber simulated spectral response for different incidence angles for the P polarisation.

Another important characteristic of these devices, that was not investigated experimentally but simulated using the same models, is the dependence of the absorption on the angle of incidence. The results of the simulations are shown in Fig. 15 and Fig. 16 for the S and P polarisations. The parametric plots demonstrate excellent performance of the mesh-absorber for large incidence angles. The absorption coefficient, averaged over the 80-400 GHz range and for both polarisations, is still above 97% at 50° ($A_{50\_S\&P} \geq 97\%$ and $R_{50\_S\&P} \leq -16$ dB) and above 87% at 70° ($A_{70\_S\&P} \geq 87\%$ and $R_{70\_S\&P} \leq -8.9$ dB). The same models also showed complete independence from the azimuthal angle.

## 7. CONCLUSIONS

We have presented the development of a novel type of broadband absorber based on mesh filter production processes. The manufacture of the first part of the device relied on the previous development of the mesh-based magnetic mirror. The second part required R&D for the evaporation of thin films of bismuth on polymers, lift-off processes for bismuth patterning and testing of the bismuth film bonding within polypropylene. These activities were run side by side with finite-element modelling (HFSS) and experimental characterisation (FTS and VNA testbeds) that fed back evaluated surface impedances and conductivities of the bismuth films between the various processes. The final device was assembled and tested in reflection at 22.5° incidence angle. The absorption coefficient, averaged across the 80-400 GHz range (a 5:1 bandwidth), resulted to be $A_{22.5\_S} = 99.2\%$ for the S polarisation, and $A_{22.5\_P} = 99.4\%$ for the P polarisation, with just a 0.2% differential absorption between the two polarisations.

The simulations showed that the device can operate efficiently across a wide range of incidence angles, with absorption still above 97% at 50° incidence and above 87% at 70° incidence.

On the thermo-mechanical side the device is: thin (750 μm, i.e. $0.6\lambda_0$ at the central frequency of 240 GHz, or $0.2\lambda$ at the lowest frequency of 80 GHz), very light (8 g for a 100 mm diameter), flexible (applicable on curved surfaces), machinable (can be cut to any shape) and suitable for cryogenic temperatures (taking into account the change of the bismuth conductivity with temperature).

We note that the broadband mesh-absorber performance stems from the magnetic mirror design reported in [30]. By keeping the same number of layers and by tuning the design parameters, the absorption can in principle be pushed at any level at the expense of reduced bandwidth. In addition, by increasing the number of layers of the graded index section (Part 1), it is possible to achieve much larger bandwidths. For example, a 5-layer device with just one more layer than the reported absorber, can achieve similar performance across a relative bandwidth of 8:1. The device also exhibits absorption at higher harmonics, although its efficiency will be limited by the period of the metal grids.

## FUNDING INFORMATION

Cardiff STFC consolidated grant (ST/N000706/1) and STFC IAA 2015.

## ACKNOWLEDGMENTS

We would like to thank Dr Matteo Ruggeri for his help during the modelling and data analysis phase.

## DISCLOSURES

The authors declare no conflicts of interest.